\pdfoutput=1
\documentclass{article}
\usepackage{spconf,amsmath,graphicx}
\usepackage{booktabs}
\usepackage{hyperref}
\usepackage{amsfonts}

\usepackage{bm}
\usepackage[numbers, sort&compress]{natbib}
\usepackage[nodisplayskipstretch]{setspace}
\hypersetup{
    colorlinks=false,
    linkcolor=black,
    filecolor=black,      
    urlcolor=black,
    pdftitle={DiffVoice: Text-to-Speech with Latent Diffusion},
    pdfpagemode=FullScreen,
}

\title{DiffVoice: Text-to-Speech with Latent Diffusion}
\name{Zhijun Liu, Yiwei Guo, Kai Yu\sthanks{Corresponding author}}
\address{MoE Key Lab of Artificial Intelligence, AI Institute\\
X-LANCE Lab, Department of Computer Science and Engineering\\
Shanghai Jiao Tong University, Shanghai, China\\
\texttt{\{zhijunliu, cantabile\_kwok, kai.yu\}@sjtu.edu.cn}
}

\begin{document}
\ninept
\maketitle
\begin{abstract}
In this work, we present DiffVoice, a novel text-to-speech model based on latent diffusion.
We propose to first encode speech signals into a phoneme-rate latent representation
with a variational autoencoder enhanced by adversarial training, and then jointly model the duration
and the latent representation with a diffusion model. Subjective evaluations on LJSpeech and LibriTTS datasets
demonstrate that our method beats the best publicly available systems in naturalness. By 
adopting recent generative inverse problem solving algorithms for diffusion models, DiffVoice achieves the state-of-the-art performance in text-based speech editing, and zero-shot adaptation. 
\end{abstract}

\section{Introduction}
\label{sec:intro}

Diffusion models (DMs) \cite{DDPM, SDE} have demonstrated great performance on image and audio generation tasks. They have also been applied to non-autoregressive text-to-speech synthesis \cite{TTSSurvey}. Most of the prior works in this direction, including \cite{DiffTTS, DiffSinger, GradTTS, GuidedTTS}, are diffusion-based acoustic models, generating log Mel spectrograms given text inputs.

Directly modeling the data density $p(\bm x_0)$ for $\bm x_0 \in \mathbb R^d$ with DMs poses several problems in application. Firstly, the intermediate latent variable $\bm x_t$ is restricted to be of the same shape as $\bm x_0$. As DM sampling requires repeated evaluation of  the score estimator $s_\theta(\bm x_t, t)$, this can be highly inefficient. Secondly, as DMs attempt to capture all modes in $p(\bm x_0)$, they tend to spend a lot of modeling capacity on imperceptible details of the data \cite{LatentDiffusion}. Latent diffusion models (LDMs) \cite{LSGM, LatentDiffusion} are proposed to alleviate these problems by first applying an encoder $f_{\phi}(\cdot)$ to encode the data into latent code $\bm z_0 = f_\phi(\bm x_0)$, then model the latent density $p_\phi(\bm z_0)$ with DMs, and finally generate data with decoder $g_\psi(\bm z_0)$.

The proposed DiffVoice model is a novel acoustic model based on LDMs. The autoencoder in DiffVoice is a VAE-GAN \cite{VAEGAN}, with dynamic down-sampling in time. It encodes a Mel-spectrogram $\bm y \in \mathbb R^{N \times D_{\text{mel}}}$ into a latent code $\bm z_0 \in \mathbb R^{M \times D_{\text{latent}}}$, where $N$ is the number of frames, $M$ is the number of phonemes. With the help of dynamic-rate down-sampling, DiffVoice can jointly model phoneme durations and Mel-spectrograms with a single diffusion model in the latent space. In contrast, prior works on diffusion acoustic models rely on additional duration predictors \cite{DiffTTS, DiffSinger, GradTTS, GuidedTTS, WaveGrad2}, and directly work on Mel-spectrograms.

DiffVoice demonstrates high performance in acoustic modeling, on both the single-speaker dataset LJSpeech\cite{LJSpeech} and the more challenging multi-speaker dataset LibriTTS \cite{LibriTTS}. As duration is jointly modeled with other factors of speech, generic inverse problem solving algorithms with DMs \cite{Posterior, Manifold} can be directly combined with \mbox{DiffVoice} for solving inverse problems in speech synthesis, including text-based speech editing, and zero-shot adaptation.

Text-based speech editing systems allow users to edit the content of recorded speech waveforms by providing the original text, and the modified text. Modifications may include insertion, deletion, and replacement of words. And the goal of such systems is to synthesis the modified part of audios with high coherence and naturalness. We demonstrate that DiffVoice can achieve state-of-the-art performance on this task, without specially tailored model designs and training procedures adopted by many prior works \cite{SpeechPainter, RetrieverTTS, CampNet, EditSpeech}. We further demonstrate that zero-shot adaptation can be solved by \mbox{DiffVoice} with state-of-the-art performance, by casting it as a speech continuation \cite{AudioLM}, or insertion problem \cite{A3T}.

Audio samples and further information are provided in the online supplement at \url{https://zjlww.github.io/diffvoice/}. We highly recommend readers to listen to the audio samples.

\vspace{-10pt}
\section{DiffVoice}
\label{sec:method}

\begin{figure*}[t!]
\centering
    \vspace{-10pt}
    \includegraphics[scale=0.85]{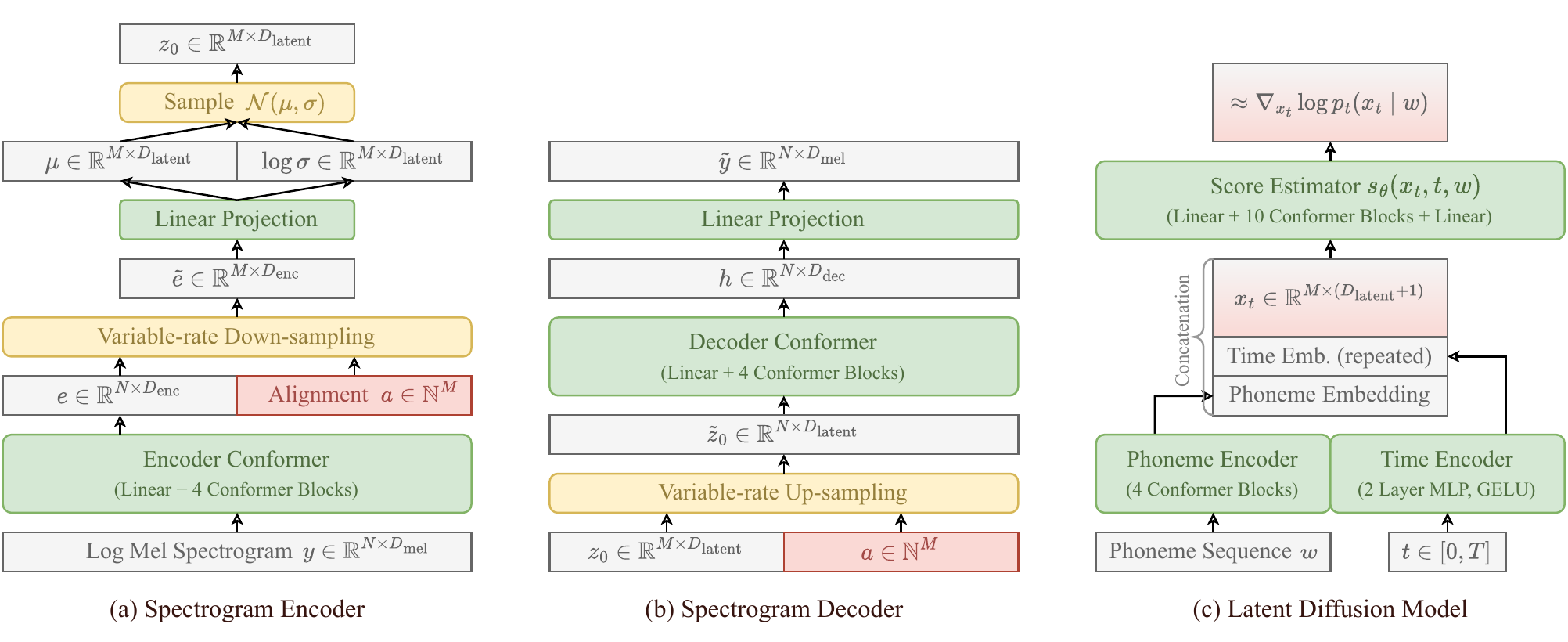}
    \vspace{-10pt}
    \caption{Architecture of the proposed DiffVoice system. $M$ is the length of phoneme sequence $w$, and $N$ is the number of frames.}
    \vspace{-10pt}
    \label{fig:main}
\end{figure*}

Suppose $\bm y \in \mathbb R^{N \times D_{\text{mel}}}$ is a log Mel spectrogram, where $N$ is the number of frames and $D_{\text{mel}}$ is the size of the Mel filter-bank. Suppose $\bm w \in \Sigma^{M}$ is the corresponding phoneme sequence where $\Sigma$ is the set of all phonemes.
\vspace{-10pt}
\subsection{Dynamic Down-Sampling of Speech}
\label{sec:vae}
DiffVoice uses a variational autoencoder \cite{VAE} to encode speech into a compact latent space. In this section, we describe the encoding and decoding of speech signals in detail.

Similar to Deep Voice \cite{DeepVoice} and TalkNet \cite{TalkNet}, we rely on a CTC based ASR model trained on phoneme sequences to obtain the alignment \cite{TTSSurvey} between $\bm w$ and $\bm y$. We use the minimal-CTC proposed in \cite{CTC} to guarantee that one and only one sharp spike is generated for each phoneme. Suppose for each $w_i$ in $\bm w = (w_i)_{i=1}^M$, its position in the CTC alignment is $a_i$ in $\bm a = (a_i)_{i=1}^M \subseteq \{1, \ldots, N\}$. Clearly $\bm a$ is strictly increasing. Let $a_0 := 0$, and $d_i := (a_i - a_{i - 1})$. Positive sequence $\bm d = (d_i)_{i=1}^M$ contains approximately the phoneme durations.

The approximate posterior $q_{\phi}(\bm{z}_0 | \bm y, \bm a)$ is defined as following (see Figure \ref{fig:main}-a). $\bm y$ is first processed by the encoder Conformer \cite{Conformer}. Then the output frame-rate latent representation $\bm e \in \mathbb R^{N \times D_{\text{enc}}}$ is down-sampled to $\tilde {\bm e} \in \mathbb R^{M \times D_{\text{enc}}}$ by gathering the values at frames $(a_i)_{i=1}^M$. $\tilde{\bm e}$ is then linearly projected and split to generate the mean $\bm{\mu} \in \mathbb R^{M \times D_{\text{latent}}}$, and log variance $\log \bm{\sigma} \in \mathbb R^{M \times D_{\text{latent}}}$. And finally, for $\bm{z}_0 \in \mathbb R^{M \times D_{\text{latent}}}$:
$$
q_{\phi}(\bm z_0 | \bm y, \bm a):= \mathcal N(\bm z_0; \bm{\mu}, \bm{\sigma})= \prod_{i=1}^M \prod_{k=1}^{D_{\text{latent}}} \mathcal N\left((z_0)_{i, k}; \mu_{i, k}, \sigma_{i, k}\right).
$$
The prior $p(\bm{z}_0)$ is defined as the standard Normal density,
$$
p(\bm z_0) := \prod_{i=1}^M \prod_{k=1}^{D_{\text{latent}}} \mathcal N\left((z_0)_{i, k}; 0, 1\right).
$$

The conditional density $p_\psi(\bm{y} | \bm{z}_0, \bm{a})$ is defined as following (see Figure \ref{fig:main}-b). $\bm z_0 \in \mathbb R^{M \times D_{\text{latent}}}$ is first up-sampled according to alignment $\bm a$ into $\tilde{\bm z}_0 \in \mathbb R^{N \times D_{\text{latent}}}$. Where $\forall 1 \le i \le M:(\tilde{z}_0)_{a_i, \cdot} = (z_0)_{i, \cdot}$ and $\forall j \notin \bm a: (\tilde{z}_0)_{j, \cdot} = \bm 0$. Then $\tilde{\bm z}_0$ is fed to the decoder Conformer to obtain $\bm h \in \mathbb R^{N \times D_{\text{dec}}}$, and then linearly projected in each frame to $\tilde{\bm{y}} \in \mathbb R^{N \times D_{\text{mel}}}$. Now we define $p_\psi$ as following, 
$$
p_\psi(\bm y | \bm z_0, \bm a) := \prod_{j=1}^N \prod_{k = 1}^{D_{\text{mel}}} \frac{1}{2b} \exp\left ({-\frac{\left |y_{j, k} - \tilde y_{j, k} \right|}{b}}\right),
$$
where $b \in (0, \infty)$ is a hyper-parameter to be tuned during training.

The variational autoencoder is trained by optimizing $\mathcal L_{\text{VAE}} = \mathbf E_{(\bm y, \bm a)} \left[\mathcal L_{\phi, \psi}(\bm y, \bm a) \right]$, where
\begin{align*}
\mathcal L_{\phi, \psi}(\bm y, \bm a) =& -\mathbf D_{\text{KL}}\left (q_\phi(\bm z_0| \bm y, \bm a) \| p(\bm z_0)\right)\\
&+ \mathbf E_{q_\phi(\bm z_0 | \bm y, \bm a)} \log p_\psi \left ( \bm y | \bm z_0, \bm a\right).
\end{align*}

\subsection{Adversarial Training}
\label{sec:adv}
\begin{figure}[h]
    \centering
    \includegraphics[scale=0.85]{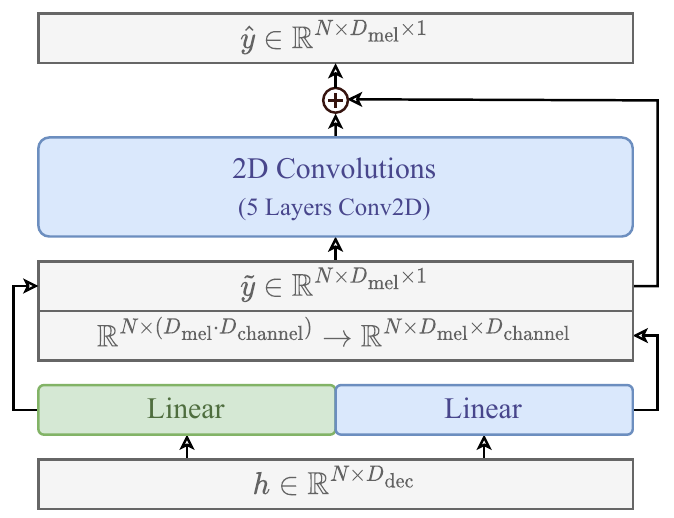}
    \vspace{-6pt}
    \caption{Modification to decoder for adversarial training. Blue blocks are added and randomly initialized in the decoder.}
    \label{fig:adv}
    \vspace{-20pt}
\end{figure}

Training with only the ELBO described in Section \ref{sec:vae} results in the autoencoder generating spectrograms lacking high-frequency details. Similar to \cite{GANSpeech, RetrieverTTS, SpeechPainter, LatentDiffusion}, we add an adversarial loss to ensure high-fidelity reconstruction.

The VAE is first trained till convergence. Then we continue training with an additional adversarial loss. In the adversarial training phase, the spectrogram decoder (Figure \ref{fig:main}-b) is extended as shown in Figure \ref{fig:adv}. A stack of randomly initialized 2D convolutions, and an extra linear projection is added to generate spectrogram residuals. The 2D convolutions are regularized by spectral norm regularization \cite{SAGAN}, and interleaved by Leaky ReLU activations. The discriminator is also a stack of 2D convolutions with spectral norm regularization, interleaved by Leaky ReLU activations.

Denote the stochastic map $(\bm y, \bm a) \mapsto \hat {\bm y}$ as the generator $\mathcal G(\bm y, \bm a)$, and $\mathcal D(\cdot)$ as the discriminator. Following \cite{GANSpeech}, we use the least-squares loss $\mathcal L_{\mathcal G}, \mathcal L_{\mathcal D}$, plus the feature matching loss $\mathcal L_{\mathcal G}^{\text{feat}}$ to train $\mathcal G$ and $\mathcal D$. The total loss during adversarial training $\mathcal L_{\text{adv}}$ is a weighted sum of $\mathcal L_{\mathcal G}, \mathcal L_{\mathcal D}, \mathcal L_{\mathcal G}^{\text{feat}}, \mathcal L_{\text{VAE}}$.
\begin{align*}
\mathcal L_{\mathcal D} &:= \mathbf E_{(\bm y, \bm a)}\left [\sum_{j, k}(\mathcal D_{j, k}(\bm y) - 1)^2\right] + \mathbf E_{(\bm y, \bm a)} \left [\sum_{j, k}\mathcal (D_{j, k}(\hat {\bm y}))^2\right]\\
\mathcal L_{\mathcal G} &:= \mathbf E_{(\bm y, \bm a)}\left [\sum_{j, k}(\mathcal D_{j, k}
(\mathcal G(\bm y, \bm a)) - 1)^2
\right ],\\
\mathcal L_{\mathcal G}^{\text{feat}} &:= \mathbf E_{(\bm y, \bm a)} \left [\frac{1}{L}\sum_{\ell=1}^L \frac{1}{d_\ell} \left \| \mathcal D^{(\ell)}(\bm y) - \mathcal D^{(\ell)}(\mathcal G(\bm y, \bm a))\right \|_1\right]. 
\end{align*}
The output of the discriminator is a 2D matrix, and $\mathcal D_{j, k}$ is the $j, k$-th value. In the feature matching loss $\mathcal L_{\mathcal G}^{\text{feat}}$, $L$ is the number of layers in $\mathcal D$. $\mathcal D^{(\ell)}$ denotes the hidden feature map at layer $\ell$ with $d_\ell$ elements.
\vspace{-5pt}
\subsection{Latent Diffusion Model}

In this section, we describe the latent diffusion model lying at the heart of \mbox{DiffVoice} (Figure \ref{fig:main}-c). After the speech autoencoder described in Section \ref{sec:vae} and \ref{sec:adv} is fully trained, we freeze its weights and use it to encode speech into latent representations.

To model the integer duration sequence $\bm d$ with a diffusion model, we first apply uniform dequantization \cite{uniform-dequantization} to $\bm d$ by sampling $\bm u \sim \operatorname{Uniform}[0, 1)^M$ and then define $\tilde {\bm d} = \bm d - \bm u$. We further take $l_j := \log(\tilde {d}_j + c_0) + c_1$, where $c_0, c_1$ are manually picked constants to normalize the distribution. Define the concatenation of $\bm l = (l_j)_{j=1}^M$ and $\bm z_0$ as $\bm x_0 := [\bm l; \bm z_0] \in \mathbb R^{M \times (D_{\text{latent}} + 1)}$. The goal of the latent diffusion model is to sample from density $p_0(\bm x_0 | \bm w)$.

We adopt the Variance Preserving SDE proposed in \cite{SDE} for generative modeling. Consider the following Itô SDE, 
\vspace{-4pt}
\begin{equation}
    \mathrm d \bm X_t =-\frac{1}{2} \beta(t) \bm X_t \mathrm d t+\sqrt{\beta(t)} \mathrm d \bm B_t,
    \label{equ:sde}
    \vspace{-4pt}
\end{equation}
$\bm X_t$ is a random process in $\mathbb R^{M \times (D + 1)}$, $t \in [0, 1]$, and $\bm B_t$ is a $\mathbb R^{M \times (D + 1)}$-valued standard Brownian motion. We used the same definition of $\beta(t)$ as in \cite{SDE}. Define $\bar \alpha(t) := \exp({- \int_0^t \beta(s) \mathrm d s})$, the transition density of Equation \ref{equ:sde} is given by
\vspace{-4pt}
$$
p_{0t}(\bm x_t | \bm x_0, \bm w) = \mathcal N(\bm x_t; \bm x_0 \sqrt{\bar \alpha(t)}, ({1 - \bar \alpha(t)}) \bm I).
\vspace{-4pt}
$$
The score estimator conditioned on text $s_\theta(\bm x_t, t, \bm w)$ is trained with denoising score matching, such that $s_\theta(\bm x_t, t, \bm w) \approx \nabla_{\bm x_t} \log p_t(\bm x_t | \bm w)$, we used the same weighting on time $\lambda_t$ as in \cite{SDE},
\vspace{-4pt}
\begin{align*}
\mathcal L_{\text{SDE}}& := \mathbf E _t\biggl\{\lambda_t \mathbf E_{(\bm x_0, \bm w)}\\
&\mathbf E _{\bm x_t|\bm x_0}\left[\left\| s_{\theta}(\bm x_t, t, \bm w)-\nabla_{\bm x_t} \log p_{0t}(\bm x_t| \bm x_0, \bm w)\right\|_2^2\right]\biggr\}.
\vspace{-8pt}
\end{align*}
During inference, first sample $\bm x_0$ from $p_0(\bm x_0 | \bm w)$ with the latent diffusion model. Then split $\bm x_0$ into $\bm l$ and $\bm z_0$. Then we reconstruct alignment $\bm a$ from $\bm l$, and decode the log Mel spectrogram $\hat {\bm y}$ from $(\bm z_0, \bm a)$ with the spectrogram decoder.
\vspace{-5pt}
\subsection{Solving Inverse Problems with DiffVoice}


Suppose $\bm o = \mathcal A(\bm{x}_0) \in \mathbb R^O$, where $\mathcal A$ is differentiable. We have
$$
\nabla_{\bm x_t} \log p_t(\bm x_t | \bm o, \bm w) = \nabla_{\bm x_t} \log p_t(\bm x_t| \bm w) + \nabla_{\bm x_t} \log p_t(\bm o | \bm x_t, \bm w).
$$
To sample from $p_0(\bm x_0 | \bm o, \bm w)$, we additionally need an estimator of $\nabla_{\bm x_t} \log p_t (\bm o | \bm x_t, \bm w)$. We found the method in \cite{Posterior, AudioInverse} performed well. Define $\pi_\theta(\bm x_t, t, \bm w)$ that approximates $\mathbf{E}\left [ \bm {x}_0 | \bm {x}_t, \bm w \right ]$,
\begin{equation}
\pi_\theta(\bm x_t, t, \bm w) := \frac{1}{\sqrt{\bar \alpha(t)}} \left ( \bm x_t + (1 - \bar \alpha(t)) s_\theta(\bm x_t, t, \bm w)\right).
\label{equ:denoising}
\end{equation}
With some weighting function $\xi(t): [0, 1] \to [0, \infty)$, we take
\begin{equation*}
\nabla_{\bm x_t} \log p_t(\bm o | \bm x_t, \bm w) \approx -\xi(t) \nabla_{\bm x_t}\lVert \mathcal A(\pi_\theta(\bm x_t, t, \bm w)) - \mathcal A(\bm x_0) \rVert^2_2.
\end{equation*}

\subsubsection{Text-Based Speech Editing and Zero-Shot Adaptation}

We only describe the algorithm for text-based speech replacement for consecutive phonemes in this section, as other forms of editing work similarly.

Given a log Mel-spectrogram $\bm y$ and the corresponding phoneme sequence $\bm w$, encode them into mean $\bm \mu$ and variance $\bm \sigma$, as described in Section \ref{sec:vae}. Split $\bm w$ into three segments $[\bm w^{(A)}; \bm w^{(B)}; \bm w^{(C)}]$, with lengths $M_A, M_B, M_C$. Replace segment $\bm w^{(B)}$ with $\bar {\bm w}^{(B)} \in \Sigma^{\bar M_B}$ to obtain $\bar {\bm w} = [\bm w^{(A)}; \bar {\bm w}^{(B)}; \bm w^{(C)}]$. The model is tasked to generate a new $\bar {\bm x}_0$ and the corresponding new spectrogram $\bar {\bm y}$, which equals to $\bm y$ except in the segment corresponding to the modification $\bar {\bm w}^{(B)}$.

With DiffVoice, text-based speech replacement can be solved in the same way as image inpainting. For ease of description, let's define a masked select function $\mathcal M$ as the map $[\bm u^{(A)}; \bm u^{(B)}; \bm u^{(C)}] \mapsto [\bm u^{(A)}; \bm u^{(C)}]$, for arbitrary matrix $\bm u$ where $\bm u^{(A)} \in \mathbb R^{M_A \times K}$ and $\bm u^{(C)} \in \mathbb R^{M_C \times K}$, for any $K \in \mathbb N^+$.

Define $\tilde {\bm \sigma} := [1; \bm{\sigma}] \in \mathbb R^{M \times (1 + D_{\text{latent}})}$, and $\tilde {\bm \mu} := [\bm l;\bm \mu] \in \mathbb R^{M \times (1 + D_{\text{latent}})}$. Let $\bm o = \mathcal M(\tilde {\bm \mu})$. And take $\nabla_{\bar {\bm x}_t} \log p_t(\bm o | \bar {\bm x}_t,\bar {\bm w})$ approximately
$$
- \xi(t)\nabla_{\bar {\bm x}_t} \left \|
\frac{
\mathcal M \left (\pi_\theta(\bar {\bm x}_t, t, \bar {\bm w})\right) - \mathcal M\left (\tilde {\bm \mu}\right )
}
{
\mathcal M \left (\tilde {\bm \sigma} \right)
}
\right \|_2^2, 
$$
where the subtraction and the division are applied element-wise. Finally, we can solve the reverse SDE or the probability-flow ODE with the modified score $\nabla_{\bar {\bm x}_t} \log p_t(\bar {\bm x}_t | \bm o, \bar{\bm w})$ to approximately sample from conditional density $p(\bar {\bm x}_t | \bm o, \bar {\bm w})$ \cite{Posterior}.

Zero-shot adaptation can be cast into an extreme form of speech insertion, where an entire new sentence is synthesized. For example, let $\bm w^{(A)}$ be the phoneme sequence of the reference speech, $\bm w^{(C)}$ the empty sequence, and $\bar {\bm w}^{(B)}$ the phoneme sequence of the new sentence. We refer to this approach as prompt-based zero-shot adaptation \cite{A3T}.

\vspace{-5pt}
\section{Experiments and Results}
\label{sec:experiment}
\vspace{-5pt}
The Conformer architecture \cite{Conformer} is widely applied in DiffVoice. The hyper-parameters for all conformer blocks in DiffVoice can be found in Table \ref{tab:conformer}. All dropout rates were set to 0.1 in our experiments. All CTC alignment models were trained on the same dataset as the acoustic models. Training and forced alignment with the minimal-CTC topology \cite{CTC} is implemented with the k2 toolkit \footnote{\url{https://github.com/k2-fsa/k2}}. More details on model training can be found in the online supplement.

\begin{table}[h!]
\centering
\begin{tabular}{@{}lllll@{}}
\toprule
\textbf{Conformer Blocks} & \textbf{Layers} & \textbf{Attention} & \textbf{Conv. Kernel} \\ \midrule
CTC model & 5 & $8 \times 64$ & 3 \\
Spectrogram Encoder & 4 & $4 \times 64$ & 13 \\
Spectrogram Decoder & 4 & $4 \times 64$ & 13 \\
Score Estimator & 10 & $8 \times 96$ & 7 \\
Speaker Encoder (\ref{sec:multispk}) & 3 & $4\times 64$ & 7 \\
Phoneme Encoder & 4 & $4 \times 128$ & 7 \\ \bottomrule
\end{tabular}
\vspace{-5pt}
\caption{Hyper-parameters of Conformer blocks in DiffVoice.}
\vspace{-10pt}
\label{tab:conformer}
\end{table}

We relied on Mean Opinion Scores (MOS) for evaluation. Listeners are asked to rate synthesized speech with scores ranging from 1.0 to 5.0 with a step of 0.5. For all evaluations in this section, 15 listeners each rated 20 randomly sampled utterances. In all evaluations, multiple stimuli with the same textual content are presented in a single trial, and the texts are presented to the listeners. All MOS scores are reported with 95\% confidence interval.

\vspace{-5pt}
\subsection{Single-Speaker Text-to-Speech}
\label{sec:synthesis}

We used LJSpeech \cite{LJSpeech} for the evaluation of text-to-speech performance on a single-speaker dataset. We leave out the same 500 sentences for testing as in VITS \cite{VITS} and GradTTS \cite{GradTTS}.

We compared our model with the best publicly available models. For VITS\cite{VITS}, and GradTTS \cite{GradTTS}, we used their official public implementation\footnote{\url{https://github.com/jaywalnut310/vits}}\footnote{\url{https://github.com/huawei-noah/Speech-Backbones}} and pretrained weights. For GradTTS the temperature was set to 1.5, and the sampler is Euler ODE sampler with 100 steps. For \mbox{DiffVoice}, we used the Euler-Maruyama sampler with 100 steps. For FastSpeech 2 \cite{FastSpeech2}, we used the implementation in ESPNet 2 \cite{ESPNet2}, where the HiFi-GAN is jointly finetuned with the FastSpeech 2 model to improve performance\footnote{\url{https://github.com/espnet/espnet}}. For GradTTS and DiffVoice we used a pretrained HiFi-GAN(v1) \footnote{\url{https://github.com/jik876/hifi-gan}} to generate waveforms. 90 sentences were synthesized for evaluation. The results can be found in Table \ref{table:ljspeech}.


\begin{table}[]
\begin{center}
\begin{tabular}{@{}lc@{}}
\toprule
\textbf{Model}          & \textbf{MOS}  \\ \midrule
GT                      & 4.81$\pm$0.04 \\
GT mel + HiFiGAN        & 4.68$\pm$0.05 \\
Auto-encoder            & 4.68$\pm$0.05 \\ \midrule
FastSpeech 2            & 4.44$\pm$0.05 \\
VITS                    & 4.54$\pm$0.05 \\
GradTTS                 & 4.01$\pm$0.06 \\
\textbf{DiffVoice}      & \textbf{4.66}$\pm$\textbf{0.04} \\ \bottomrule
\end{tabular}
\caption{Results for speech synthesis on LJSpeech.}
\label{table:ljspeech}
\end{center}
\vspace{-20pt}
\end{table}

\vspace{-5pt}
\subsection{Multi-Speaker Text-to-Speech}
\label{sec:multispk}

We used the union of the ``train-clean-100'' and ``train-clean-360'' splits in LibriTTS \cite{LibriTTS}, for the evaluation of text-to-speech performance on a multi-speaker dataset. It contains about 245 hours of speech from 1151 speakers. We randomly selected 500 utterences for evaluation. All audios were down-sampled to 16kHz.

We extended DiffVoice with a speaker encoder jointly trained with the score estimator, referred to as DiffVoice (Encoder) in the following. The speaker encoder is a Conformer with input $\bm x_0$. The speaker embedding is obtained by mean pooling the output of the Conformer. The embedding is repeated in time and concatenated with the other inputs to the score estimator. For VITS and FastSpeech 2, we used their multi-speaker extensions in ESPNet2 \cite{ESPNet2} conditioning on X-vectors. For YourTTS \cite{YourTTS} and Meta-StyleSpeech \cite{MetaStyle} we used their official public implementation and pre-trained weights. For FastSpeech 2, Meta-StyleSpeech, and DiffVoice, we synthesized waveforms with a universal HiFi-GAN trained on ``train-clean-460''.

For evaluation of text-to-speech performance in the multi-speaker setting. We report MOS, Similarity MOS (Sim-MOS), and Speaker Encoder Cosine Similarity (SECS) following \cite{YourTTS}. SECS scores are computed with the speaker encoder in Resemblyzer \footnote{\url{https://github.com/resemble-ai/Resemblyzer}}.

For seen speakers, we randomly selected 40 sentences from 40 different speakers in ``train-clean-460''. We used speaker-level X-vectors for FastSpeech 2 and VITS, and random reference audios for the DiffVoice (Encoder) model. The evaluation results can be found in Table \ref{table:libritts}. Note that we used the ground truth audios as references for similarity tests. 

For the evaluation of zero-shot adaptation, we used the same 21 reference audios as YourTTS \cite{YourTTS}. 5 sentences were synthesized per reference. We used utterance-level X-vectors for X-vector models in the zero-shot evaluation. DiffVoice (Prompt) is a system using prompt-based zero-shot adaptation described in Section \ref{sec:editing}. To sample from this model, we used the Euler-Maruyama sampler with 300 steps.

\begin{table}[]
\begin{center}
\begin{tabular}{@{}lccc@{}}
\toprule
\textbf{Model}               & \textbf{MOS}& \textbf{Sim-MOS}   & \textbf{SECS} \\ \midrule
GT                           & 4.81$\pm$0.03 & N/A & N/A \\
GT mel + HiFi-GAN            & 4.50$\pm$0.04 & 4.95$\pm$0.02 & 0.992 \\
Auto-encoder                 & 4.52$\pm$0.04 & 4.94$\pm$0.02 & 0.985 \\ \midrule
FastSpeech 2 (X-vector)      & 3.18$\pm$0.04 & 3.74$\pm$0.06 & 0.896 \\
VITS (X-vector)              & 3.80$\pm$0.05 & 4.10$\pm$0.05 & 0.919 \\
\textbf{DiffVoice (Encoder)} & \textbf{4.32}$\pm$\textbf{0.04} & \textbf{4.72}$\pm$\textbf{0.04} & \textbf{0.938} \\ \bottomrule
\end{tabular}
\caption{Results for speech synthesis with seen speakers on LibriTTS.}
\label{table:libritts}
\end{center}
\vspace{-10pt}
\end{table}

\begin{table}[]
\begin{center}
\begin{tabular}{@{}lccc@{}}
\toprule
\textbf{Model}             & \textbf{MOS}   & \textbf{Sim-MOS}  & \textbf{SECS} \\ \midrule
GT                         & 4.89$\pm$0.03  & 4.94$\pm$0.02 &  0.870      \\
GT mel + HiFi-GAN          & 4.75$\pm$0.03  & 4.92$\pm$0.02 &  0.862      \\
Auto-encoder               & 4.76$\pm$0.04  & 4.83$\pm$0.04 &  0.857      \\ \midrule
Meta-StyleSpeech           & 2.82$\pm$0.07  & 3.15$\pm$0.08 &  0.764      \\
YourTTS                    & 3.46$\pm$0.06  & 3.51$\pm$0.06 &  0.806      \\
FastSpeech 2 (X-vector)    & 3.42$\pm$0.06  & 3.44$\pm$0.07 &  0.752      \\
VITS (X-vector)            & 4.14$\pm$0.06  & 3.80$\pm$0.06 &  0.800      \\
DiffVoice (Encoder)        & 4.32$\pm$0.07  & 3.52$\pm$0.07 &  0.703      \\ 
\textbf{DiffVoice (Prompt)}& \textbf{4.66}$\pm$\textbf{0.05}  & \textbf{4.61}$\pm$\textbf{0.04} &  \textbf{0.854}      \\ \bottomrule
\end{tabular}
\caption{Results for zero-shot adaptation on LibriTTS.}
\label{tab:zeroshot}
\end{center}
\vspace{-20pt}
\end{table}

\subsection{Text-based Speech Editing}
\label{sec:editing}

We evaluate the performance of text-based speech inpainting, which is a special case of replacement, by comparing with samples from {RetrieverTTS} \cite{RetrieverTTS}. We used the same SDE sampler as in DiffVoice (Prompt). We strictly followed the evaluation set up described in \cite{RetrieverTTS}. The MOS for three different mask durations can be found in Table \ref{table:inpainting}. Please refer to \cite{RetrieverTTS} for further details.

\begin{table}[h]
\begin{center}
\begin{tabular}{@{}lccc@{}}
\toprule
\textbf{Model}  & \textbf{MOS@short}& \textbf{MOS@mid}  & \textbf{MOS@long} \\ \midrule
GT              & - & 4.90$\pm$0.04 & - \\
RetrieverTTS    & 4.13$\pm$0.09 & 3.56$\pm$0.11 & 3.56$\pm$0.08 \\
\textbf{DiffVoice} & \textbf{4.43}$\pm$\textbf{0.07} & \textbf{4.09}$\pm$\textbf{0.08} & \textbf{4.08}$\pm$\textbf{0.08} \\ \bottomrule
\end{tabular}
\caption{Results for text-based speech inpainting.}
\label{table:inpainting}
\end{center}
\vspace{-20pt}
\end{table}

\vspace{-5pt}
\section{Conclusions and Future Works}
DiffVoice demonstrated strong performance on text-to-speech synthesis, speech editing, and voice cloning. But sampling remains relatively slow, requiring hundreds of neural function evaluations. Better SDE sampler, and other acceleration methods for diffusion models might decrease the sampling time while maintaining the same sample quality. Using better intermediate representations other than log Mel spectrograms, and applying improved techniques for waveform generation will also improve the performance.
\vspace{-5pt}
\section{Acknowledgements}

This study was supported by State Key Laboratory of Media Convergence Production Technology and Systems Project (No. SKLMCPTS2020003) and Shanghai Municipal Science and Technology Major Project (2021SHZDZX0102).

\vfill\pagebreak


\typeout{}
\bibliographystyle{IEEEbib}
\bibliography{refs}

\end{document}